\documentclass[aps,prb,reprint,groupedaddress]{revtex4-1}

\usepackage{graphicx}
\usepackage{color}
\usepackage{enumerate}
\usepackage[normalem]{ulem}

\begin{document}


\title{Thermal stability and irreversibility of skyrmion-lattice phases in Cu$_2$OSeO$_3$}


\author{Koya Makino}
\affiliation{Institute of Multidisciplinary Research for Advanced Materials, Tohoku University, 2-1-1 Katahira, Sendai 980-8577, Japan}

\author{Johannes D. Reim}
\email[]{j.reim@tagen.tohoku.ac.jp}
\affiliation{Institute of Multidisciplinary Research for Advanced Materials, Tohoku University, 2-1-1 Katahira, Sendai 980-8577, Japan}

\author{Daiki Higashi}
\affiliation{Institute of Multidisciplinary Research for Advanced Materials, Tohoku University, 2-1-1 Katahira, Sendai 980-8577, Japan}

\author{Daisuke Okuyama}
\affiliation{Institute of Multidisciplinary Research for Advanced Materials, Tohoku University, 2-1-1 Katahira, Sendai 980-8577, Japan}

\author{Taku J. Sato}

\affiliation{Institute of Multidisciplinary Research for Advanced Materials, Tohoku University, 2-1-1 Katahira, Sendai 980-8577, Japan}

\author{Yusuke Nambu}
\affiliation{Institute for Materials Research, Tohoku University, Sendai 980-8577, Japan}

\author{Elliot P. Gilbert}
\affiliation{Australian Centre for Neutron Scattering, Australian Nuclear Science and Technology Organization, Kirrawee DC, New South Wales 2232, Australia}

\author{Norman Booth}
\affiliation{Australian Centre for Neutron Scattering, Australian Nuclear Science and Technology Organization, Kirrawee DC, New South Wales 2232, Australia}

\author{Shinichiro Seki}
\affiliation{RIKEN Center for Emergent Matter Science (CEMS), Wako, Saitama 351-0198, Japan}
\affiliation{PRESTO, Japan Science and Technology Agency (JST), Tokyo 102-8666, Japan}

\author{Yoshinori Tokura}
\affiliation{RIKEN Center for Emergent Matter Science (CEMS), Wako, Saitama 351-0198, Japan}
\affiliation{Department of Applied Physics and Quantum Phase Electronics Center (QPEC), University of Tokyo, Tokyo 113-8656, Japan}

\date{\today}

\begin{abstract}
Small angle neutron scattering measurements have been performed to study the thermodynamic stability of  skyrmion-lattice phases in Cu$_2$OSeO$_3$.
We found that the two distinct skyrmion-lattice phases [SkX(1) and SkX(2) phases] can be stabilized through different thermal histories;  by cooling from the paramagnetic phase under finite magnetic field, the SkX(2) phase is selected.
On the other hand, the 30$^{\circ}$-rotated SkX(1) phase becomes dominant by heating the sample from the ordered conical phase under finite field.
This difference in stabilization is surprisingly similar to the irreversibility observed in spin glasses.
The zero-field cooling results in the co-existence of the two phases.
It is further found that once one of the skyrmion-lattice phases is formed, it is hardly destabilized.
This indicates unusual thermal stability of the two skyrmion-lattice phases originating from an unexpectedly large energy barrier between them.
\end{abstract}


\maketitle


\section{Introduction}
Topological spin textures have attracted growing interest in the condensed matter physics community.
One representative example of such spin textures is a magnetic skyrmion in chiral magnets - a swirling spin structure carrying a topological quantum number~\cite{Bogdanov89,Nagaosa13}.
The magnetic skyrmion, condensed in a triangular lattice arrangement, was first observed in the chiral helimagnet MnSi in 2009~\cite{MuhlbauerS09}.
Since then, skyrmion-lattice phases have been reported in a number of compounds, such as FeGe~\cite{Yu11}, (Fe,Co)Si~\cite{Munzer10}, Cu$_2$OSeO$_3$~\cite{Seki12,Adams12}, Co$_8$Zn$_8$Mn$_4$~\cite{TokunagaY14}, GaV$_4$S$_8$~\cite{KezsmarkiI15}, and so on.
Among them, Cu$_2$OSeO$_3$ attracts special attention because it is a rare example of an insulating, and consequently multiferroic, compound hosting the skyrmion lattice.
The formation of the skyrmion lattice in this compound was confirmed by small angle neutron scattering (SANS)~\cite{Seki12,Adams12} and Lorentz microscopy~\cite{Seki12_1}, whereas its multiferroicity was investigated in detail by macroscopic measurements~\cite{Seki12_1,Seki12_2}.
The response of the skyrmion lattice to the external electric field has also been studied using the SANS technique, where electric-field-induced rotation of the skyrmion lattice has been observed~\cite{White12,White14}.
Recent studies further indicate intriguing phase transformations in the thin film~\cite{Rajeswari15} or surface~\cite{Zhang16} of Cu$_2$OSeO$_3$, suggesting the importance of the geometrical confinement for the skyrmion phase stability.
Through these extensive studies, the formation of the skyrmion-lattice phase and its multiferroic response have been firmly confirmed.
In addition, theoretical investigations revealed Dzyaloshinskii-Moriya interactions to be key for inducing the helical and skyrmion state \cite{Yang2012}.

Nonetheless, there is an unsolved puzzle. Contradicting temperature-field ($T$-$H$) phase diagrams for the bulk Cu$_2$OSeO$_3$ sample in the earlier studies,
Seki {\it et al.} reported that there are two skyrmion-lattice phases, {\it i.e.}, the skyrmion-lattice 1 [SkX(1)] and skyrmion-lattice 2 [SkX(1)] phases, which are related by a 30$^{\circ}$ rotation with each other~\cite{Seki12}.
The two phases are adjacent in the $T$-$H$ phase diagram, where SkX(1) [SkX(2)] is stabilized at higher (lower) external field.
On the other hand, Adams {\it et al.}~\cite{Adams12} reported only the SkX(2) phase, covering the entire skyrmion-lattice region in the phase diagram.

In this study, we have reinvestigated the $T$-$H$ phase diagram of Cu$_2$OSeO$_3$ using the SANS technique.
Careful attention was paid to the temperature and magnetic-field history, through which the skyrmion-lattice phases are stabilized.
We have clearly shown that the different skyrmion-lattice phases are stabilized under different $T$-$H$ protocols; field-cooling (FC) stabilizes the SkX(2) phase, whereas the SkX(1) phase tends to form after heating up from the conical phase under finite magnetic field.
The zero-field-cooling (ZFC) results in the co-existence of those two phases.
Surprisingly, once either one of the skyrmion-lattice phases is stabilized, the phase does not relax into the other phase within the present experimental time scale.
This was confirmed even in the vicinity of the higher-temperature phase boundary, where the thermal fluctuations should be maximized.
This unexpected robustness of each skyrmion-lattice phase suggests an abrupt formation of an energy barrier of considerable height immediately after entering into the coexistence region.

\section{Experimental}
A single crystal of Cu$_2$OSeO$_3$ with the approximate dimensions 8\,mm $\times$ 5\,mm $\times$ 3\,mm ($\sim 330$\,mg) was grown by the chemical vapor transport method~\cite{Miller10}.
SANS experiments were performed using the QUOKKA instrument\cite{Gilbert2006} installed at the OPAL reactor of Australian Nuclear Science and Technology Organization.
Incident neutrons with the wavelength $\lambda \sim 5$\,\AA\space are selected using a neutron velocity selector, with the wavelength distribution ${\rm d}\lambda/\lambda \sim 10$\%.
Supplemental high $Q$ resolution measurements were performed with several longer wavelengths.
The Cu$_2$OSeO$_3$ crystal was placed on a sapphire sample mount with its [110] axis parallel to the incident neutron beam $\vec{k}_{\rm i}$, and with its [001] axis along the horizontal direction.
It may be noted that to gain higher scattering intensity enabling detailed phase diagram investigation, the grown crystal was used without cutting into a regular shape.
The sample mount was then set to the cold head of a closed-cycle $^4$He refrigerator, loaded in a 5\,T horizontal field superconducting magnet with the magnetic field parallel to $\vec{k}_{\rm i}$.
The sample temperature was monitored by a temperature sensor placed on the sapphire sample mount.
A circular Cd aperture with diameter $d \sim 5$\,mm was placed in the vicinity of the sample so that neutrons do not irradiate the edges of the sample.
Background scattering was estimated from the SANS patterns measured at the paramagnetic temperature 60\,K, and was subtracted from all the data reported here.
For comparison, Fig. \ref{fig8} shows selected scattering patterns before background subtraction normalized only to the measurement time and detector sensitivity.

\section{Results and discussion}

\subsection{Phase stabilization for different $T$-$H$ histories}

\begin{figure}
\includegraphics[scale=0.85, angle=0, trim={0cm 0cm 0cm 0cm}]{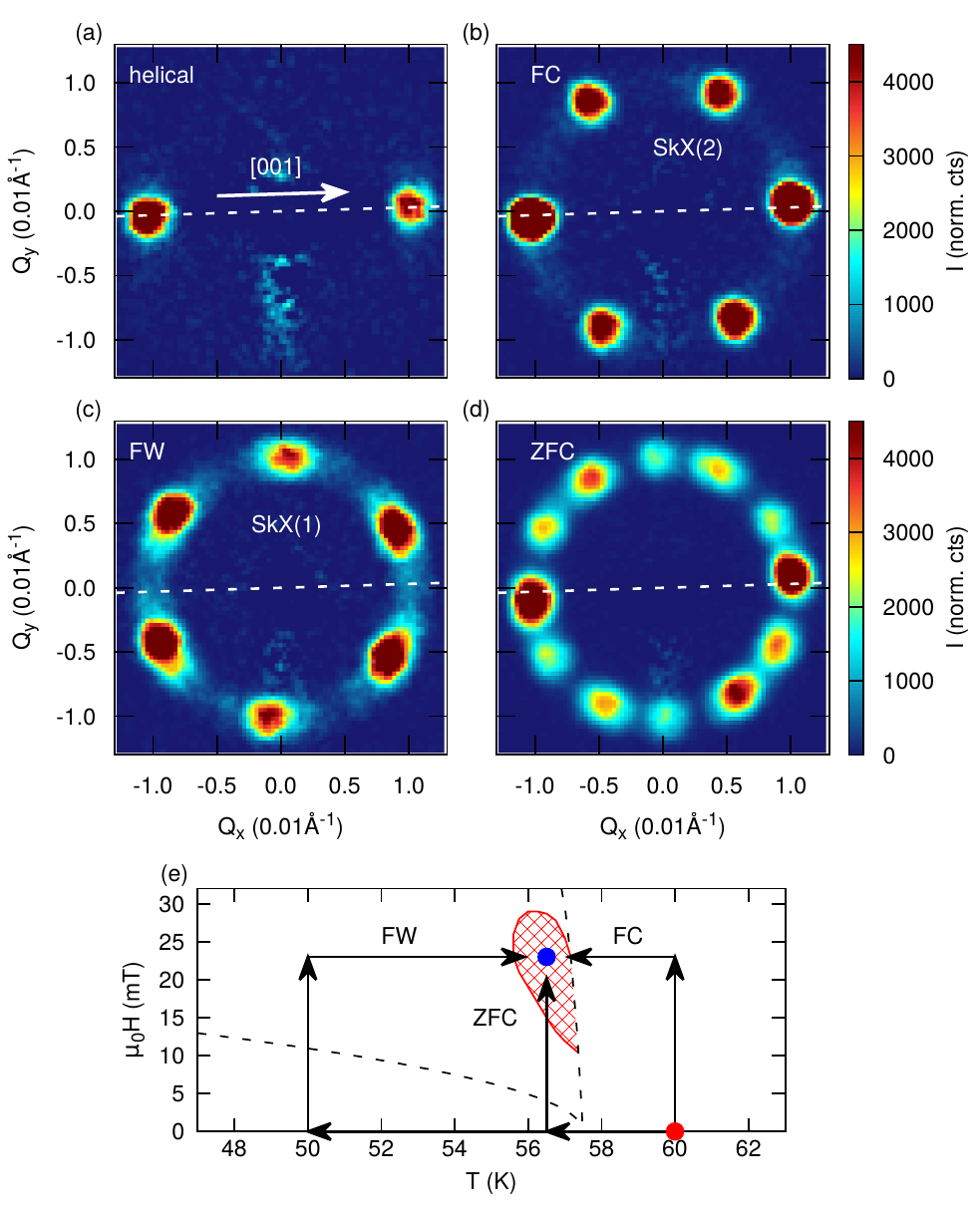}
\caption{(Color online) (a) SANS pattern measured at $\mu_0 H = 0$\,mT and $T = 56.5$\,K, where the helical phase is formed.
The crystallographic [001] direction was deduced from the Bragg positions of the helical phase, which are denoted by the dashed lines in Figures (a-d).
The intensity shown in (a) is scaled by a factor of 2 to increase visibility.
(b-d) SANS patterns measured at $\mu_0 H = 23$\,mT and $T = 56.5$\,K using the three different temperature-field ($T$-$H$) protocols: (b) field-cooling (FC), exemplary for the SkX(2) scattering pattern; (c) field-warming (FW), exemplary for the SkX(1) scattering pattern; (d) zero-field-cooling (ZFC).
The intensity is capped according to the color scale.
(e) Schematic illustration of the three $T$-$H$ protocols used for the SANS measurements shown above.
The details of the $T$-$H$ protocols are given in the main text.
}\label{fig1}
\end{figure}

The formation of the zero-field helical phase was first confirmed at $T = 56.5$\,K.
The resulting SANS pattern is shown in Fig.~\ref{fig1}(a).
Clearly, the expected two Bragg reflections were observed at $Q = 0.01$\,\AA$^{-1}$, in agreement with the earlier reports.
A slight rotation of the Bragg positions is due to the misalignment of the crystal.
The crystallographic [001] direction is determined from the Bragg positions, and is indicated by the dashed line.
Further temperature scans at zero-field provide evidence for the helical phase transition at around $T=57.25$\,K (see Fig.~\ref{fig10}).

For the measurements under finite magnetic field, the following three different protocols were used in the present study.
The first one is the FC protocol, where the sample is initially heated up to the paramagnetic temperature ($T = 60$\,K), and then cooled down to the target temperature with the rate 2\,K/min under the finite magnetic field.
The second one is the field-warming (FW) protocol, where the sample is heated up from the helical/conical phase ($T = 50$\,K) under the finite magnetic field.
The third one is the ZFC protocol, where the sample is initially heated up to the paramagnetic temperature ($T = 60$\,K), then cooled down to the target temperature at the rate of 2\,K/min under zero external magnetic field and, after the temperature has stabilized, the external magnetic field is applied.
The three protocols are schematically illustrated in Fig.~\ref{fig1}(e).

The SANS pattern obtained at $T = 56.5$\,K and $\mu_0 H = 23$\,mT using the FC protocol is shown in Fig.~\ref{fig1}(b).
The figure shows well-defined six-fold Bragg reflections, being a typical SANS pattern for the skyrmion-lattice phase.
Based on these data alone, this observation cannot be discerned from a multi-domain state; however, a skyrmion-lattice was recently reported in Cu$_2$OSeO$_3$ thin films using cryo-Lorentz transmission electron microscopy~\cite{Rajeswari15}.
Two horizontal peaks appear along the [001] direction, indicating that this skyrmion-lattice phase corresponds to the SkX(2) phase reported earlier.
It may be further noted that scattering intensity is higher for the two Bragg peaks along the [001] direction, compared to the other four peaks.
This may be due to a slight misalignment of the crystal and/or misalignment of the internal magnetic field due to the irregular shape of the used crystal {\color{red}(less than $5^{\circ}$)}, and is seen for all the SANS patterns obtained in this study.
However, this does not affect the discussion nor conclusions of this report.
For the FW protocol, completely different SANS patterns were observed.
The SANS pattern obtained at $T = 56.5$\,K and $\mu_0 H = 23$\,mT with the FW protocol is shown in Fig.~\ref{fig1}(c).
Again the six-fold Bragg reflections were observed, however, the reflection positions are rotated by approximately 30$^{\circ}$ compared to those in the FC measurement.
This SANS pattern corresponds to the SkX(1) phase in the earlier report~\cite{Seki12}.
It may be noted that in the SkX(1) phase, the Bragg peaks are broadened along the azimuthal direction, and a ring-like diffuse contribution may be seen in between the Bragg peaks.
The SANS pattern at $T = 56.5$\,K and $\mu_0 H = 23$\,mT obtained with the ZFC protocol is shown in Fig.~\ref{fig1}(d).
With the ZFC protocol, clearly the two  skyrmion-lattice phases coexist, resulting in the two six-fold diffraction patterns distributed nearly symmetrically. 
This can be also seen in the azimuthal variation displayed in Fig.~\ref{fig9}.
From the above results, we can conclude that stabilization of the skyrmion-lattice phases deterministically depends on the $T$-$H$ history, and a different skyrmion-lattice phase [either SkX(1) or SkX(2)] can be selectively stabilized by selecting a different $T$-$H$ protocol.

\subsection{Phase diagrams}

\begin{figure}
\includegraphics[scale=0.8, angle=0, trim={0cm 0cm 0cm 0cm}]{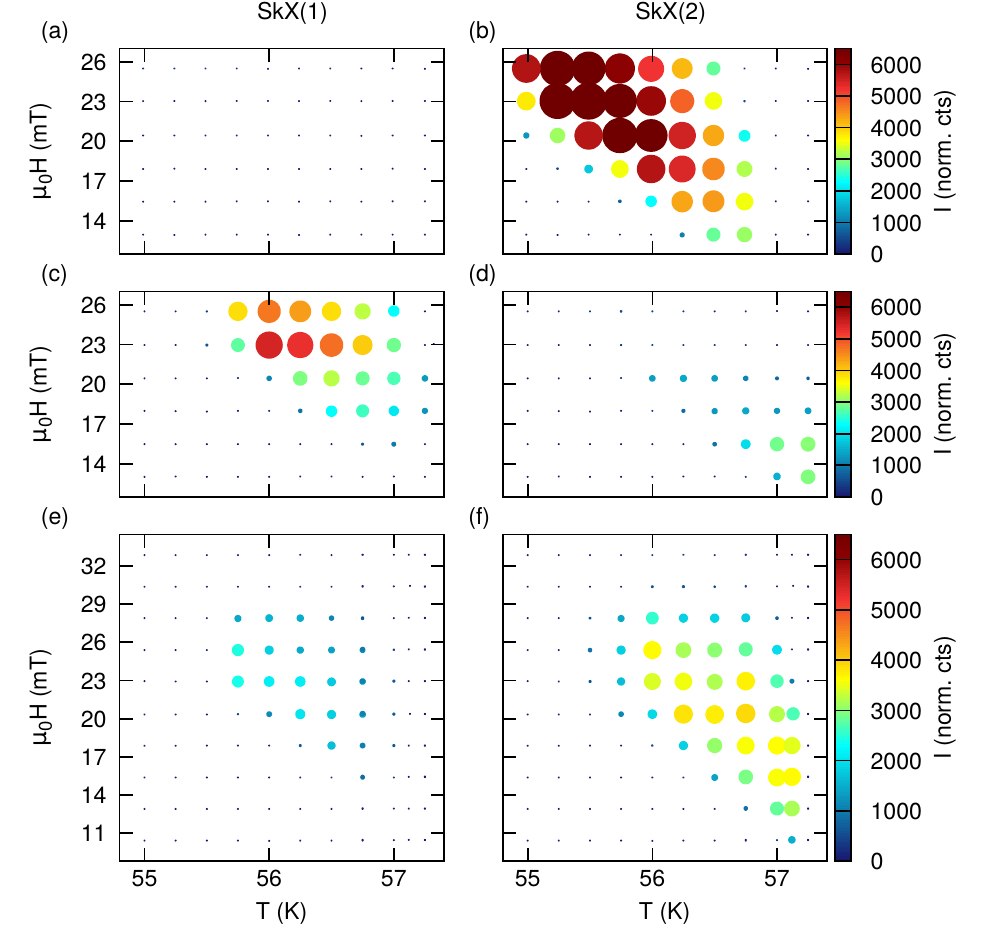}
\caption{(Color online) Two dimensional intensity maps for the SkX(1) (left column) and SkX(2) (right column) components, obtained by the fitting to the model function Eq.~(\ref{eq:phidepmodel}).
Top panels, middle panels, and bottom panels represent the results obtained using the FC, FW, and ZFC protocols.
The position of the dots denotes the measured temperature and field points, whereas their color and size represents the intensity of each component.
The minimum sized dots indicate the positions of measurements exhibiting a respective intensity of less than 50\,normalised counts.
Details of the temperature-field sequence are given in Appendix~\ref{appendixB}.
}\label{fig2}
\end{figure}

Next, to obtain (quasi-)equilibrium phase diagrams for the different $T$-$H$ histories, we measured the SANS patterns in wider $T$- and $H$-ranges using the three distinct $T$-$H$ protocols.
Specifically, we set the sample temperature and field to those of the first measurement point using the FC, FW, and ZFC protocols, and then measured a number of SANS patterns by sequentially changing $T$ and $H$ with the steps of $\Delta T = 0.25$\,K and $\mu_0 \Delta H = 2.5$\,mT.
(Finer steps were used whenever necessary.)
The details of the $T$-$H$ sequences are described in Appendix~\ref{appendixB}.
For a quantitative analysis, the azimuthal variation at $Q=0.0100(24)$\AA$^{-1}$ in each SANS pattern is fitted using the Eq. \ref{eq:phidepmodel} (see Appendix~\ref{appendixC} for the complete description) which allows to obtain the
integrated intensity and rotation angles of the SkX(1) and SkX(2) phases, as well as the intensity of the ring-like diffuse scattering component. 

The integrated intensity of the Bragg peaks for the SkX(1) and SkX(2) phases, obtained with the FC protocol, is shown in the two-dimensional maps, Figs.~\ref{fig2}(a) and \ref{fig2}(b), respectively.
In the entire $T$- and $H$-ranges, the SkX(2) intensity dominates with almost negligible contribution from the SkX(1) phase.
This clearly indicates that the SkX(2) phase is formed in the entire skyrmion-lattice region in the $T$-$H$ phase diagram when using the FC protocol.
It should be noted that this single-phase behavior of the skyrmion-lattice phase is consistent with the earlier report~\cite{Adams12}.

The integrated intensity maps obtained with the FW protocol are given in Figs.~\ref{fig2}(c) and \ref{fig2}(d) for the SkX(1) and SkX(2) phases, respectively.
In contrast to the FC result, the SkX(1) phase is predominantly formed in a wide $T$-$H$ range, in particular in the lower-$T$ and higher-$H$ region.
Nonetheless, faint scattering intensity can also be observed for SkX(2); however, these peaks exhibit an irregular symmetry.
In order to maintain consistency with the other protocols a regular model was used for fitting SkX(2).
With increasing $T$ and decreasing $H$, the SkX(1) intensity gradually decreases, and instead the SkX(2) intensity grows.
In the vicinity of the paramagnetic phase boundary ($T = 57.25$\,K), we found that the SkX(2) phase is uniquely stabilized.
It may be noted that the phase boundary to the paramagnetic phase seems to shift slightly to the higher temperature, compared to that observed with the FC protocol.
While this hysteresis behavior could be partly due to an insufficient temperature equilibration time for the FC runs shown in Figs.~\ref{fig2}(a) and \ref{fig2}(b); it should be noted that temperature steps were quite small $\Delta T = 0.25$\,K; longer equilibration times were employed throughout the studies at the first measuring temperature (at least 600\,s).
An alternative possibility is that this hysteresis behavior may be related to the first-order nature for the skyrmion-paramagnetic phase transition, observed earlier in dynamical susceptibility measurements~\cite{LevaticI14}.

\begin{figure}
\includegraphics[scale=0.9, angle=0, trim={0cm 0cm 0cm 0cm}]{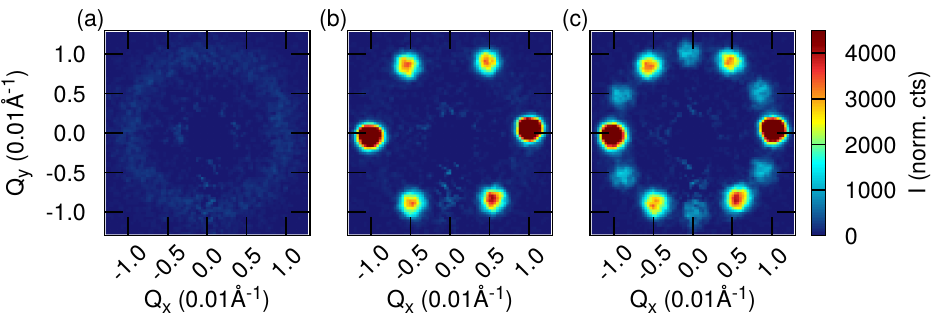}
\caption{(Color online) SANS patterns measured at $\mu_0 H = 17.9$\,mT and (a) $T = 57.25$, (b) 57.125, and (c) 57.0\,K using the ZFC protocols.
At 57.25\,K, the system is apparently in the paramagnetic phase, whereas single SkX(2) phase was confirmed at 57.125\,K.
The mixture of SkX(1) and SkX(2) was observed at 57.0\,K, indicating that SkX(2) is single-phase in a very narrow temperature range under the ZFC condition.
}\label{fig3}
\end{figure}

The integrated intensity maps obtained with the ZFC protocol are given in Figs.~\ref{fig2}(e) and \ref{fig2}(f) for the SkX(1) and SkX(2) peaks.
For almost the entire $T$-$H$ range, the SkX(1) and SkX(2) peaks were simultaneously observed, indicating that the two phases coexist under the ZFC protocol.
For the lower-$T$ side, the SkX(1) phase is relatively dominant, whereas at higher-$T$ the SkX(2) phase contributes significantly.
Again, in a very narrow $T$-region at the higher-$T$ end of the skyrmion-lattice phase, the SkX(2) phase becomes the only observed phase.
This behavior is qualitatively similar to the phase diagram reported by Seki {\it et al.}~\cite{Seki12}.
However, we do not observe here a complete switching of the skyrmion phases from SkX(2) to SkX(1) with increasing the magnetic field; this highlights the significance of small differences in the experimental setup as the same sample was used in both experiments.
The stability of the SkX(2) phase in the vicinity of the paramagnetic phase boundary was further confirmed by measuring SANS patterns at three temperatures $T = 57.25$, 57.125 and 57.0\,K using the ZFC protocol.
The resulting SANS patterns are shown in Figs.~\ref{fig3}(a-c).
At $T = 57.25$\,K, the system is apparently paramagnetic with weak ring-shape scattering, {\color{red}which is} indicative of the development of critical fluctuations and similar to those observed in MnSi in the vicinity of its $T_{\rm c}$~\cite{GrigorievSV14}.
Upon cooling to $T = 57.125$\,K, a clear six-fold SkX(2) pattern was observed, indicating the single phase nature at this temperature.
This diffraction pattern changes into the superposition of SkX(2) and SkX(1) phases at $T = 57.0$\,K.
This result clearly shows that the SkX(2) phase is single-phase only in a very narrow temperature range.

\subsection{Skyrmion-lattice deformation at lower temperatures}

\begin{figure}
\includegraphics[scale=0.9, angle=0, trim={0cm 0cm 0cm 0cm}]{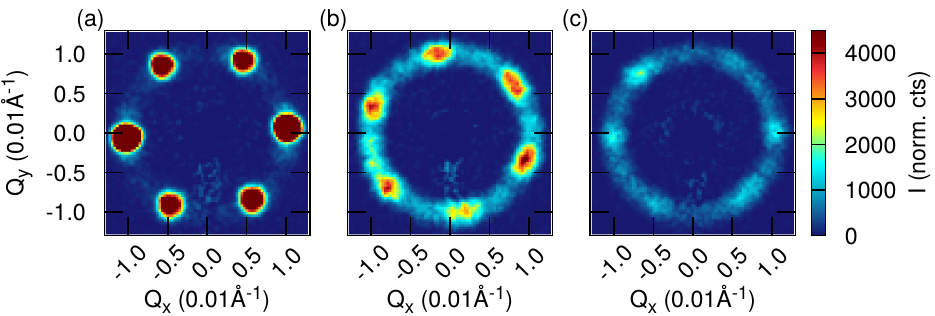}
\caption{(Color online) SANS patterns measured at $\mu_0 H = 22.9$\,mT and $T = 55.5$\,K using the (a) FC, (b) FW, and (c) ZFC protocols.
Clearly, six sharp Bragg peaks were observed in the (a) FC panel, whereas the peaks are considerably broadened in the (b) FW result.
For the ZFC panel, Bragg peaks become much weaker, and instead the diffuse ring becomes dominant.
}\label{fig4}
\end{figure}

\begin{figure}
\includegraphics[scale=0.8, angle=0, trim={0cm 0cm 0cm 0cm}]{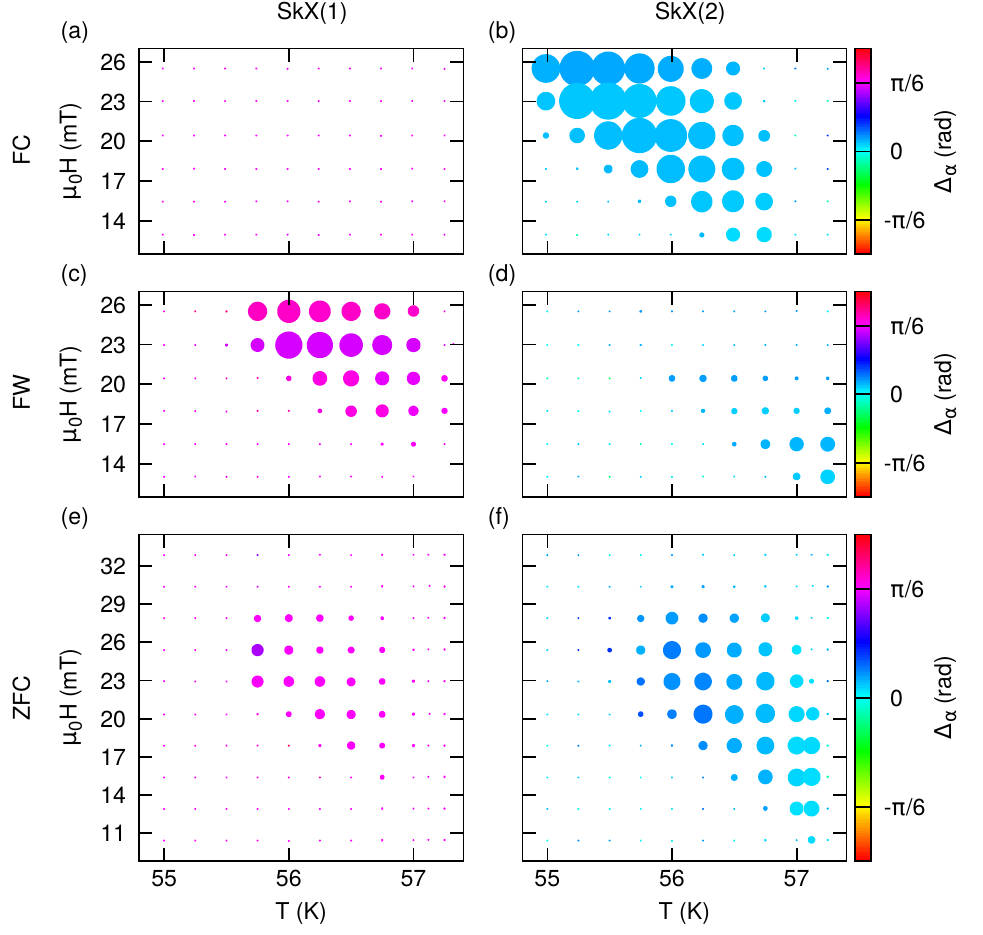}
\caption{(Color online) Two dimensional maps for the angle parameters $\Delta_1$ (a, c, e) and $\Delta_2$ (b, d, f) for the SkX(1) and SkX(2) components, obtained by the fitting of the model function Eq.~(\ref{eq:phidepmodel}).
Top panels, middle panels, and bottom panels represent the results obtained using the FC, FW, and ZFC protocols.
The color of the dots represent the angle values with the size indicating the intensity given in Fig.~\ref{fig2}.
Details of the temperature-field sequence are given in Appendix~\ref{appendixB}.
}\label{fig5}
\end{figure}

\begin{figure}
\includegraphics[scale=1.0, angle=0, trim={2cm 0cm 1cm 0cm}]{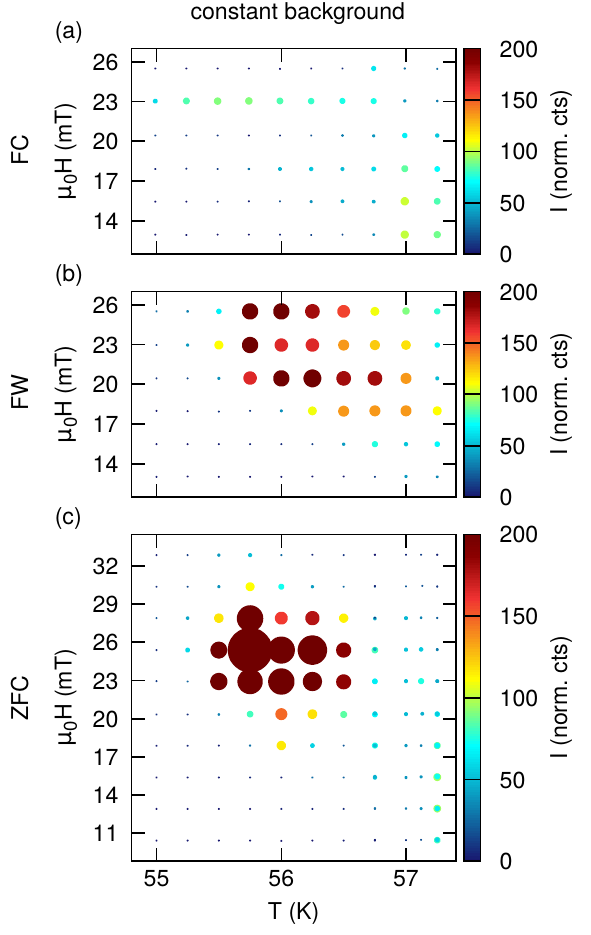}\ 
\caption{(Color online) $T$- and $H$-dependence of the constant term $C$ in Eq.~(\ref{eq:phidepmodel}) as an estimate of the intensity of the ring-like diffuse scattering, represented by the dots' color and size.
Results obtained with the SANS patterns measured under (a) FC, (b) FW, (c) ZFC protocols are shown.
}\label{fig6}
\end{figure}

In contrast to the stabilization of the single SkX(2) phase at the higher-$T$ phase boundary, an intriguing $T$-$H$-history dependence was observed at lower temperatures, close to the lower boundary of the skyrmion-lattice phase.
Figures~\ref{fig4}(a), \ref{fig4}(b), and \ref{fig4}(c) show the SANS patterns at $T = 55.5$\,K obtained with the FC, FW, and ZFC protocols.
The clear SkX(2) pattern was observed only for the FC protocol.
For the FW protocol, although six-fold reflections were visible, they are rotated significantly however from the expected SkX(1) peak positions, being almost in the middle of the SkX(1) and SkX(2) positions.
Furthermore, the Bragg peaks are considerably broadened in the azimuthal direction, accompanied by a weak SkX(2) component.
These results indicate that the SkX(1) phase stabilized by the FW protocol is rotated globally from the ideal angle with a considerable angle distribution, and coexists with a minor SkX(2) phase.
In the ZFC result, the hexagonal Bragg peaks are much suppressed, and instead a ring-like diffuse scattering becomes dominant, as shown in Fig.~\ref{fig4}(c).
This indicates significant local disorder in the skyrmion-lattice structure formed under the ZFC protocol.
Such a ring-like feature in the SANS pattern was also observed in the skyrmion-lattice phase of MnSi near its lower temperature phase boundary under the ZFC protocol~\cite{GrigorievSV14}.
Nonetheless, it should be emphasized that in this study we show that the local (as well as global) disorder (rotation) strongly depends on the $T$-$H$ history through which the skyrmion-lattice phase is stabilized.

To obtain insight into the global rotation of the skyrmion lattices and its $T$- and $H$-dependence, the angle parameters, $\Delta_1$ and $\Delta_2$, obtained in the fitting to Eq.~(\ref{eq:phidepmodel}), are shown in Fig.~\ref{fig5}.
$\Delta_1$ ($\Delta_2$) is the angle between the crystallographic [001] direction (cf. Fig. \ref{fig1}) and the nearest skyrmion-lattice peak of the SkX(1) [SkX(2)] phase (see Eq. B2 for mathematical definition).
For the FC protocol [Figs.~\ref{fig5}(a) and \ref{fig5}(b)], the SkX(2) peaks appear at almost ideal angles in the entire $T$-$H$ region with no apparent $T$-dependence.
The SkX(1) component is very weak, so that the angle parameter was only obtained around $H \sim 26$\,mT, and was constant with $T$.
For the FW protocol, both SkX(1) and SkX(2) angles were found to be $T$- and $H$-dependent, as shown in Figs.~\ref{fig5}(c) and \ref{fig5}(d).
It may be noteworthy that the SkX(2) angle converges into the ideal $\Delta_2 = 0$ at higher-$T$ and lower-$H$ phase boundary around $T \sim 57$\,K and $\mu_0 H \sim 14$\,mT.
The ZFC result also shows finite $T$- and $H$-dependence for the SkX(1) angles at lower-$T$ and higher-$H$ regions, whereas the SkX(2) angle converges into the ideal one at higher-$T$ boundary.
Summarizing all the behaviors for the three different temperature protocols, the skyrmion-lattice phase formed from the paramagnetic phase is largely aligned with underlying crystal lattice, and can remain aligned even at lower temperatures.
In contrast, the skyrmion-lattice phases formed across the lower-$T$ or lower-$H$ phase boundary are rotated from the ideal angles; the rotation angle depends on temperature and/or field.
This suggests a strong pinning effect in the lower temperature region, as well as a very flat free energy landscape in the azimuthal direction.
We note here that in the earlier theoretical analysis, such an anisotropy to fix the skyrmion-lattice in the azimuthal direction is indeed expected to be weak, as it originates from the sixth or higher order terms in free energy~\cite{White14}.

Next, we discuss the ring-like diffuse scattering component, which is related to the disorder in the skyrmion lattices at a much shorter length scale.
The ring-like diffuse scattering component may be captured by the term $C$ of the function $BG(\phi)$ in Eq.~(\ref{eq:phidepmodel}) used for the least-square fitting (see Appendix~\ref{appendixC}).
In Figs.~\ref{fig6}(a), \ref{fig6}(b), and \ref{fig6}(c), we plot $C$ as a function of $T$ and $H$ for the FC, FW and ZFC cases, respectively.
For the FC case, the diffuse contribution is very weak over the whole $T$-$H$ range.
This again indicates the robustness of the SkX(2) phase stabilized through FC.
On the other hand, the diffuse scattering clearly appears for the FW and ZFC cases.
Nonetheless, the diffuse scattering appears in different $T$-$H$ regions for FW and ZFC.
For FW, the diffuse ring appears in a $T$-$H$ region where the SkX(1) phase transforms into the SkX(2) phase.
This naturally suggests that the local deformation is necessary to the phase transformation.
In contrast, for ZFC, the diffuse ring appears in a slightly lower-$T$ region, and seems to have less relation to the SkX(1) to SkX(2) phase transformation.
Recent Lorentz-transmission-electron-microscopy (LTEM) study shows that skyrmions form a glassy structure in vicinity of the lower $H$ phase boundary, and related dislocations remain even in the long-range-ordered skyrmion-lattice phase~\cite{Rajeswari15}.
This was confirmed in even more recent magnetisation measurements~\cite{Qian16}.
Hence, we speculate that such glassy-structure-related local dislocations exist for the skyrmion-lattice structure formed from the lower-$H$ side through ZFC.
It should be noted, however, that the LTEM result is obtained with a thin (150\,nm) sample where the skyrmion-lattice stability is largely enhanced.
Further study with thicker samples is necessary for a conclusive discussion.

\subsection{Time relaxation}

\begin{figure}
\includegraphics[scale=1.0, angle=0, trim={0cm 0cm 0cm 0cm}]{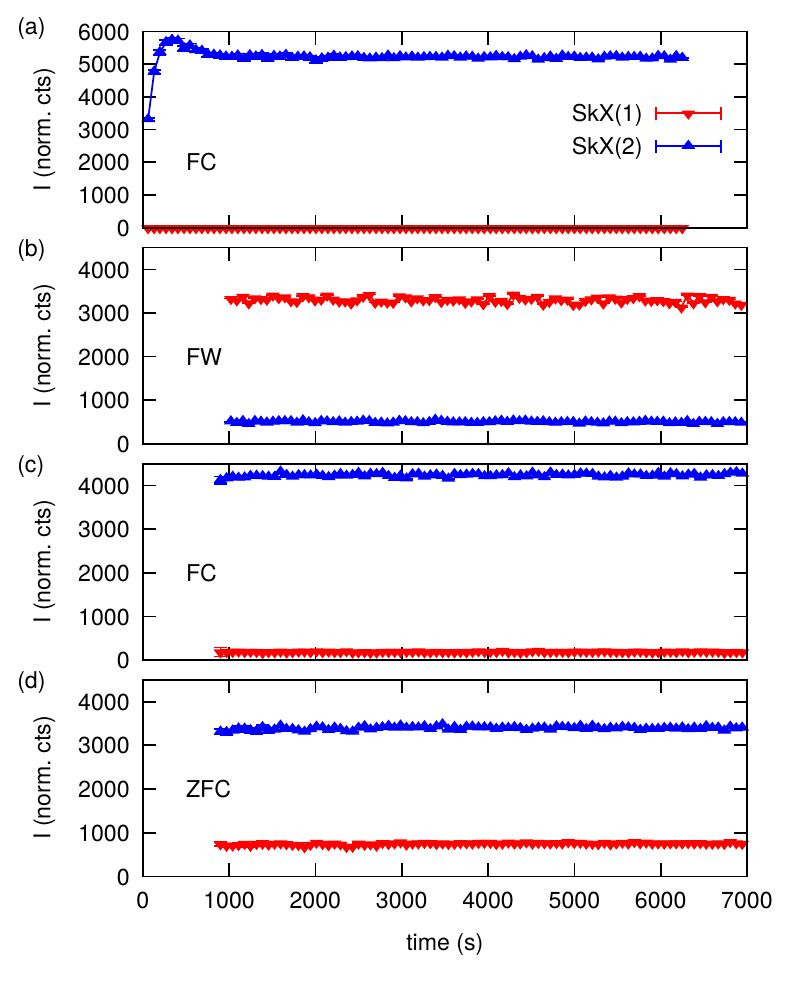}
\caption{(Color online) Time evolutions of integrated intensity for the SkX(1) and SkX(2) peaks.
(a and b) time evolution measured at $\mu_0 H = 17.4$\,mT and $T = 56.5$\,K using the (a) FC and (b) FW protocols.
(c and d) time evolution measured at $\mu_0 H = 20.4$\,mT and $T = 57$\,K using the (c) FC and (d) ZFC protocols.
The intensity change in the first 600\,s period in (a) is due to the temperature equilibration between the sample and sample holder (sapphire plate) to which the temperature sensor is attached.
}\label{fig7}
\end{figure}

As the SkX(1) and SkX(2) phases are found to be stabilized under the different $T$-$H$ histories, it is possible that one of the two phases is only metastable, and there is a relaxation from the metastable phase to the globally stable phase.
To check this possibility, we studied the temporal relaxation of the SkX(1) and SkX(2) phase formed with the FC, FW, and ZFC protocols.
Figures~\ref{fig7}(a) and (b) show the temporal evolution of the integrated intensity for the SkX(1) and SkX(2) phases obtained at $T = 56.5$\,K and $\mu_0 H = 17.4$\,mT using the FC and FW protocols, respectively, whereas those in Figs.~\ref{fig7}(c) and (d) are at $T = 57$\,K and $\mu_0 H = 20.4$\,mT obtained under the FC and ZFC protocols.
We found that temperature stabilization requires at most 600\,s in our current experimental setup; this was confirmed by observing the intensity evolution of the helical Bragg peaks after changing temperature.
We, hence, waited for approximately 1000\,s for the temperature stabilization before measuring the time relaxation.
One exception is the case shown in Fig.~\ref{fig7}(a), where the data acquisition started 60\,s after the temperature of the sapphire mount reached the target temperature, and hence strong intensity variation was seen in the initial 600\,s period.
In the FC result at $T = 56.5$\,K, SkX(2) dominates all the temporal region, whereas the SkX(1) intensity is negligible.
Neither the SkX(2) nor SkX(1) intensity changed on our experimental time scale, indicating that no relaxation occurs from the SkX(2) to SkX(1) phase once the former is formed.
At the same temperature, under the FW protocol, the SkX(1) phase becomes predominantly formed with the minority SkX(2) phase.
Again, the intensity ratio between the SkX(1) and SkX(2) phases remains constant over the experimental time scale; no relaxation was observed from the SkX(1) to SkX(2) phase.

At the higher temperature $T = 57$\,K, which is only 0.125\,K below the temperature where SkX(2) becomes globally stable, one may expect that if one of the skyrmion phases were metastable the free energy barrier between the metastable and stable phases would be significantly reduced.
Nonetheless, our observations in Figs.~\ref{fig7}(c) and \ref{fig7}(d) clearly show that neither SkX(1) nor SkX(2) grows.
This certainly indicates that once one of the SkX(1) and SkX(2) phases is formed, it hardly destabilizes into the other phase even in the vicinity of the phase boundary.
The energy barrier between the two skyrmion-lattice phases, consequently, has to be considerably larger than the thermal energy $T \sim 57$\,K.
It is surprising to see an abrupt formation of such a tall energy barrier with decreasing temperature by only 0.125\,K.
At the present moment we are not aware of any mechanism explaining the robustness of the two phases, and merely speculate here that it may be related to the topological nature of the skyrmion-lattice phases.
Further study is necessary to elucidate this phase stability issue.

\section{Conclusions}

We have performed SANS experiments on the $T$-$H$ phase diagram of the insulating skyrmion-lattice compound Cu$_2$OSeO$_3$, with special attention being paid to the $T$-$H$ history through which the skyrmion-lattice phase is stabilized.
Under the same experimental conditions a phase diagram with the single SkX(2) phase was obtained for the FC protocol, whereas the SkX(1) phase was predominantly observed for the FW protocol.
Comparing the behavior for FC and FW, the clear separation exhibits similarities to the irreversibility observed in spin glasses.
Two phase coexistence was confirmed in the ZFC phase diagram.
We have shown that the phase diagram reported by Adams {\it et al.}~\cite{Adams12} can be reproduced using the FC protocol, while the ZFC protocols leads to a phase diagram similar to the one by Seki {\it et al.}~\cite{Seki12}.
However, the presented protocol dependent stabilization is inconsistent with the one reported by White {\it et al.}~\cite{White12}; this may be related to the difference in the direction of the applied magnetic field and sample shape.
It was further found that once one of the  skyrmion-lattice phases is formed, it becomes unexpectedly stable, and no relaxation to the other phase is observed.
This is a very striking feature, suggesting the abrupt formation of an unexpectedly tall energy barrier between the two skyrmion-lattice phases immediately below the upper phase boundary.

\begin{acknowledgments}
The authors thank N. Nagaosa for stimulating discussions.
This work was partly supported by Grants-In-Aid for Scientific Research (24224009, 15H05458, 15H05883, 16H04007, and 16K13842) from MEXT of Japan.
Travel expenses for the experiment on QUOKKA at ANSTO was partly sponsored by the General User Program of ISSP-NSL, University of Tokyo.
Work at IMRAM was partly supported by the Research Program "Dynamic Alliance for Open Innovation Bridging Human, Environment and Materials". 
\end{acknowledgments}

\appendix
\section{Raw scattering pattern}\label{appendixA}
\begin{figure}[h]
\includegraphics[scale=0.85, angle=0, trim={0cm 0cm 0cm 0cm}]{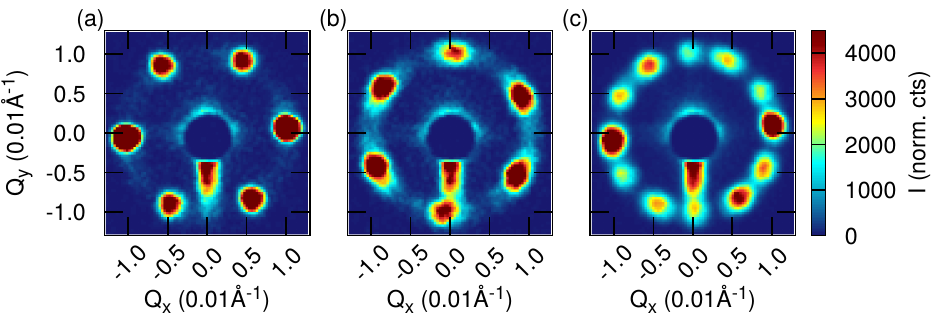}
\caption{(Color online) SANS patterns before background subtraction measured at $\mu_0 H = 23$\,mT and $T = 56.5$\,K using the three different temperature-field ($T$-$H$) protocols: (a) field-cooling (FC); (b) field-warming (FW); (c) zero-field-cooling (ZFC).
}\label{fig8}
\end{figure}

\renewcommand{\thesubsection}{\alph{subsection}}
The background scattering was determined in the paramagnetic regime at 60\,K.
A strong contribution is present close to the origin of the reciprocal space due to the direct beam and along the vertical scattering axis.
While the first feature diminished rather quickly, the latter one even protrudes up the Q-positions of the skyrmion Bragg peaks. 
To minimize its influence, the normalized background scattering was subtracted from scattering patterns prior to analysis.
For clarity, one may see the influence of the background scattering by comparing the raw scattering pattern in Fig.~\ref{fig8} with Fig.~\ref{fig1}. 

\section{Details of $T$-$H$ protocols for phase diagram study}\label{appendixB}

In this appendix, we describe the $T$-$H$ protocols which were used to obtain the three phase diagrams shown in Figs.~\ref{fig2}, \ref{fig5}, and \ref{fig6}.

\subsection{FC phase diagram}
\noindent
(i)~The sample is heated up to the paramagnetic temperature ($T = 60$\,K);
(ii)~Target magnetic field is applied to the sample;
(iii)~The sample is cooled down with the rate of 2\,K/min to the first measurement temperature $T = 57.25$\,K;
(iv)~After waiting 600\,s for the temperature stabilization, the SANS pattern is recorded for 1\,min;
(v)~The temperature is lowered with the decrement of $\Delta T = 0.25$\,K, and then the SANS pattern is recorded for 1\,min;
There was no waiting time for this FC protocol, due to the beamtime restriction;
Repeat this step until the temperature reaches the lowest measurement temperature ($T = 55$\,K);
(vi)~Repeat the above steps [from (i)] with the target magnetic field increased with the increment $\mu_0 \Delta H = 2.5$\,mT, until the maximum target field is reached.

\subsection{FW phase diagram}
\noindent
(i)~The sample is cooled down to $T = 50$\,K, where the helical/conical phase is formed;
(ii)~Target magnetic field is applied to the sample;
(iii)~The sample is heated up with the rate of 2\,K/min to the first measurement temperature $T = 55$\,K;
(iv)~After waiting 720\,s for the temperature stabilization, the SANS diffraction pattern is recorded for 1\,min;
(v)~The temperature is increased with an increment of $\Delta T = 0.25$\,K with 120\,s waiting for temperature stabilization, and then the SANS pattern is recorded for 1\,min;
Repeat this step until the temperature reaches the highest measurement temperature ($T = 57.25$\,K);
(vi)~Set the magnetic field to zero and repeat the above steps [from (i)] with the target magnetic field increased with the increment $\mu_0 \Delta H = 2.5$\,mT, until the maximum target field is reached.

\subsection{ZFC phase diagram}
\noindent
(i)~The sample is heated up to the paramagnetic temperature ($T = 60$\,K);
(ii)~The sample is cooled down to the target measurement temperature, the first one being $T = 57.25$\,K, with the rate of 2\,K/min under zero magnetic field;
(iii)~Wait for relatively short time (60\,s) for ZFC due to beamtime restriction;
(It is expected that this shorter period does not significantly change the obtained phase diagram since, while measuring the lower-field part, $0 \leq \mu_0 H \leq 12.5$\,mT, the sample reaches its equilibrium temperature target value.)
(iv)~Magnetic field is set to the target magnetic field, and the SANS pattern was recorded for 1 min. 
After recording, the magnetic field is once turned off ($\mu_0 H = 0$), and then set to the next target field increased with the increment of $\mu_0 \Delta H = 2.5$\,mT.
This step is repeated until the maximum field is reached;
(v)~The magnetic field is turned off ($\mu_0 H = 0$);
(vi)~Repeat the above steps [from (i)] with the target temperature decreased with the decrement 0.25\,K, until the lowest measurement temperature is reached.


\section{Analysis details for quantifying SANS patterns}\label{appendixC}
\begin{figure}
\includegraphics[scale=0.85, angle=0, trim={0cm 0cm 0cm 0cm}]{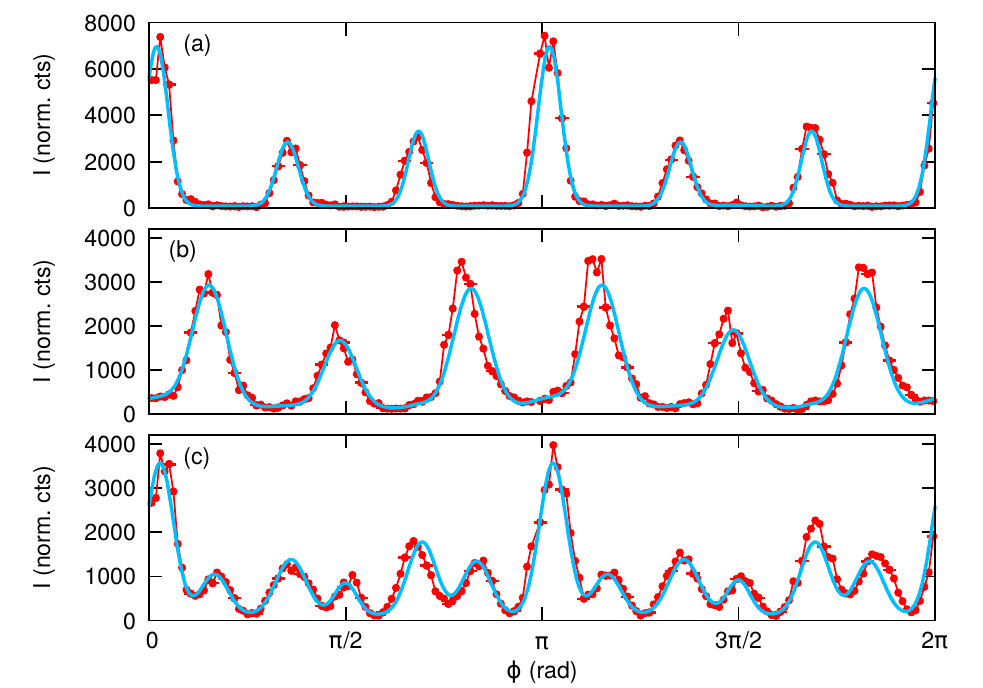}
\caption{(Color online) Azimuthal angle dependence of the diffraction intensity measured at $\mu_0 H = 23$\,mT and $T = 56.5$\,K using the FC (a), FW (b) and ZFC protocol (c).
The data shown in Fig.~\ref{fig1}(b-d) is integrated in the radial $|\vec{Q}|$-range $0.0076 \leq |\vec{Q}| \leq 0.0124$\,\AA$^{-1}$.
Blue line is the result of fitting the respective data set to the model function Eq.~(\ref{eq:phidepmodel}).
Red dots correspond to the observed data with the intervals indicating binning widths.
Representative error bars are given for a few selected data points.
}\label{fig9}
\end{figure}

The obtained SANS patterns were analyzed using the following procedure.
First, the azimuthal angle dependence of the scattering intensity was obtained by integrating the two-dimensional map in the range $0.0076 \leq |\vec{Q}| \leq 0.0124$\,\AA$^{-1}$.
The representative example for the azimuthal angle dependence obtained from the 2D data shown in Fig.~\ref{fig1}(d) is shown in Fig.~\ref{fig9}.
The azimuthal angle dependence is subsequently fitted to the following model function:
\begin{equation}\label{eq:phidepmodel}
I(\phi) = I_{\rm SkX(1)}(\phi) + I_{\rm SkX(2)}(\phi) + BG(\phi),
\end{equation}
where
\begin{eqnarray}
I_{{\rm SkX}\alpha}(\phi) &=& \sum_{n = 0}^{2} \left [ G(\phi; \Delta_{\alpha} + \frac{\pi}{3} n, I_{\alpha n}, \sigma_{\alpha n}) \right. \nonumber \\
  & + & \left. G(\phi; \Delta_{\alpha}  + \frac{\pi}{3} n + \pi, I_{\alpha n}, \sigma_{\alpha n}) \right ],\\
BG(\phi) & = & C + G(\phi; \frac{3}{2}\pi, I_{\rm BG}, \sigma_{\rm BG}).
\end{eqnarray}
The $I_{\rm SkX(1)}(\phi)$ and $I_{\rm SkX(2)}(\phi)$ terms stand for the intensity from the SkX(1) and SkX(2) phases, respectively. 
$BG(\phi)$ consists of the $\phi$-independent intensity $C$ and a weak feature localized at $\phi = 3\pi/2$.
The latter is commonly observed in the obtained SANS patterns as exemplified in Fig.~\ref{fig8}, and is possibly attributed to incomplete background subtraction.
For the peak profile, we use a Gaussian function $G(\phi; \phi_0, I_0, \sigma) = I_0/(\sqrt{2\pi}\sigma) \exp[-(\phi - \phi_0)^2/2\sigma^2]$.
In the fitting procedure, the parameters $\Delta_{\alpha}, I_{\alpha n}, \sigma_{\alpha n}, C, I_{\rm BG}$ and $\sigma_{\rm BG}$ are optimized, with the soft constraints $\Delta_{1} \simeq \pi/6$ [for the SkX(1) phase], and $\Delta_{2} \simeq 0$ [for the SkX(2) phase].
Finally, the $\phi$-integrated intensity for the SkX$\alpha$ contribution, $\int_0^{2\pi} {\rm d}\phi I_{{\rm SkX}\alpha}$, is obtained.

\section{Helical phase transition}\label{appendixD}
The helical phase in Cu$_2$OSeO$_3$ is present at low magnetic fields and is separate from the skyrmion phase.
As the stabilization of the skyrmion phases shows a strong protocol dependence, the actual history of each measurement is relevant for the comprehension.
Therefore, the transition temperature of the helical phase was determined in zero-field using a cooling and heating temperature scan.
The hysteresis behavior between the two scans can at least partly be attributed to an insufficient temperature equilibration time.
Nonetheless, both reveal a phase transition at $T_C\sim57.25$\,K.
\begin{figure}[h]
\includegraphics[scale=0.85, angle=0, trim={0cm 0cm 0cm 0cm}]{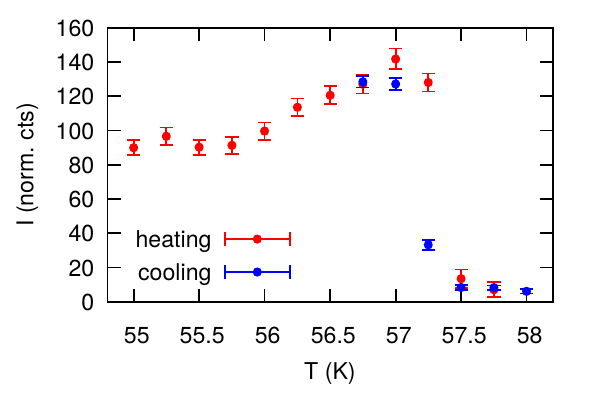}
\caption{(Color online) Integrated scattering intensities determined from the helical peaks at zero magnetic field. 
Both heating and cooling scans reveal a phase transition at around $T=57.25$\,K.
}\label{fig10}
\end{figure}

%

\end{document}